\documentclass{article}
\usepackage{graphicx}
\usepackage{amsmath}
\usepackage{cite}

\title{Satellite Internet of Things Research Report}
\author{Zhang Shutao}

\begin{document}
	
	\maketitle
	
	\section{Research Overview}
	
	This section introduces satellite IoT from three perspectives: background introduction, demand analysis, and the development status at home and abroad \cite{zhang2020efficient, yu2017iterative, ning2022multi, li2022real, zhang2023statistical, zhang2023physics}.
	
	\subsection{Background Introduction}
	
	Satellite communication is a method of communication between two or more earth stations using artificial earth satellites as relay stations. It has advantages such as wide coverage, large communication capacity, good durability, and strong adaptability, making it an important means to achieve seamless global communication. The world's first artificial satellite was launched by the Soviet Union in October 1957, which sent signals back to Earth. The first synchronous communication satellite was launched by NASA in July 1963, but it was not yet a geostationary satellite. The initial semi-experimental, semi-practical geostationary satellite "Early Bird" was launched in April 1965 for commercial satellite communication between Europe and America, marking the practical stage of satellite communication.
	
	Satellite communication systems include all equipment for communication and ensuring communication, generally consisting of four parts: space subsystem, communication earth station, tracking telemetry command subsystem, and monitoring management subsystem, as shown in Figure 1. The main body of the space subsystem is the communication satellite, which mainly includes one or more transponders, each capable of simultaneously receiving and retransmitting signals from multiple earth stations, thus serving as a relay station. The communication earth station is a microwave radio transmission and reception station, through which users access satellite lines for communication. The tracking telemetry command subsystem is responsible for tracking and measuring the satellite to ensure it accurately enters the designated position in the geostationary orbit. After the satellite is in normal operation, it needs to periodically correct its orbital position and maintain its attitude. The monitoring management subsystem is responsible for detecting and controlling the communication performance of the fixed satellite before and after the service is activated, such as monitoring the basic communication parameters like satellite transponder power, satellite antenna gain, RF frequency, and bandwidth to ensure normal communication.
	
	Currently, the frequency bands used by satellites mainly include C-band, Ku-band, and Ka-band. According to type, satellites can be divided into scientific satellites, experimental satellites, and application satellites. According to application, satellites can be classified into communication satellites, meteorological satellites, reconnaissance satellites, navigation satellites, and astronomical satellites. Based on orbit altitude, satellites can be classified into the following three categories:
	
	\begin{figure}[h]
		\centering
		\includegraphics[width=0.9\textwidth]{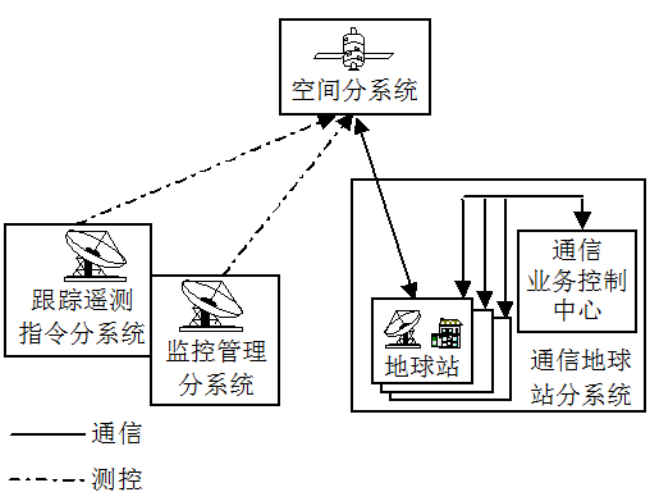}
		\caption{Basic Composition of Satellite Communication System}
		\label{fig:satellite_communication_system}
	\end{figure}
	
	\begin{itemize}
		\item \textbf{Low Earth Orbit (LEO) Satellites}: Orbit altitude between 700 and 1500 km, with an orbital period of 2 to 4 hours. LEO satellites have advantages such as reduced propagation delay and lower transmission loss. A constellation of multiple LEO satellites can achieve seamless global coverage.
		\item \textbf{Middle Earth Orbit (MEO) Satellites}: Orbit altitude between 8000 and 20000 km, with an orbital period of 4 to 12 hours.
		\item \textbf{High Earth Orbit (HEO) Satellites}: Orbit altitude greater than 20000 km, including geostationary satellites at about 35786 km, with an orbital period of one sidereal day, i.e., 23 hours, 56 minutes, and 4 seconds. When the orbital inclination of the geostationary satellite relative to the equatorial plane is zero, it is called a geostationary orbit (GEO).
	\end{itemize}
	
	Since its first proposal at MIT in 1999, the IoT has received significant attention. Traditional IoT can be divided into three layers: the perception layer, the network layer, and the application layer, transmitted through local area network methods such as WIFI, Bluetooth, and Zigbee. With the widespread application and continuous maturation of IoT technology, Low Power Wide Area Network (LPWAN) has been proposed, including technologies like Extended Coverage-GSM (EC-GSM) operating in licensed bands, Narrowband IoT (NB-IoT), and LoRa, Sigfox in unlicensed bands, aiming to provide services for IoT applications requiring long-distance transmission, low power consumption, and massive connectivity. The 3GPP has stated that NB-IoT will continue to evolve to support massive Machine Type Communications (mMTC) scenarios in 5G, aiding the development of 5G IoT.
	
	The International Telecommunication Union (ITU) proposed four application scenarios for satellite and 5G integration: cell backhaul, relay to station, communication on the move, and hybrid multicast. The 3rd Generation Partnership Project (3GPP) has proposed corresponding protocols for the satellite-ground integration architecture in Release 15 (R15), Release 16 (R16), and Release 17 (R17) \cite{3gpp.38.811, 3gpp.22.822, 3gpp.22.261}. The SaT5G Alliance focuses on researching communication technologies for satellite and 5G integration and demonstrated its validation platform at the European Conference on Networks and Communications in 2018.
	
	\subsection{Demand Analysis}
	
	Traditional IoT already supports a variety of applications, such as smart agriculture, smart home, and smart transportation. However, due to the diverse terrain on Earth, the construction of base stations for ground communication networks is limited by natural environments and geographical locations, making it difficult to cover areas like deserts, forests, and oceans. Satellite communication networks can complement ground communication networks in terms of coverage. Satellites are distributed in the sky and are not limited by geographical locations. A well-designed satellite constellation can provide all-weather, seamless global coverage and, compared to ground communication networks, offers higher reliability and resilience in cases of earthquakes, floods, and other disasters.
	
	Therefore, the so-called satellite IoT is the IoT that uses satellite communication networks for information transmission. Satellite IoT can achieve wide-area coverage, long-distance transmission, and information collection in remote areas, playing a crucial role in fields such as field exploration, marine environment monitoring, and global emergency search and rescue. In 2015, Liu Gang and others proposed the space-based IoT, and with continuous research, the architecture and construction ideas of the space-based IoT have been proposed.
	
	\subsection{Development Status at Home and Abroad}
	
	Although the application of satellite communication in IoT has not yet been explicitly proposed domestically or internationally, many systems fall under the category of satellite IoT. Examples include foreign systems like Orbcomm, Argo, Iridium, Globalstar, and domestic systems like Beidou, Hongyan, and Xingyun.
	
	The Orbcomm satellite system, built by Orbital Sciences Corporation of the USA and Teleglobe of Canada, started commercial operation in November 1998. It includes 36 satellites, each weighing 43 kg. Users can collect remote data, monitor vehicle locations, and send and receive short messages through this system. The Argo satellite system, established by the French National Centre for Space Studies, NASA, and the National Oceanic and Atmospheric Administration, has an orbital altitude of 850 km and is mainly used for real-time global ocean monitoring. The Iridium satellite system, launched by Iridium Communications, consists of 66 satellites, each weighing around 670 kg, providing burst data services globally, mainly for the US military. The Globalstar satellite system, initiated by Loral Corporation and Qualcomm, includes 48 low-orbit satellites with an orbital altitude of about 1414 km, offering seamless satellite mobile communication services to users. Additionally, some satellite IoT systems are under construction, such as HeliosWire of Canada, planning to launch 30 satellites to support data collection from 5 billion sensors, and Fleet Space Technologies of Australia, aiming to build a nano-satellite constellation to help global IoT devices achieve low-bandwidth connections, with the first batch of 1 million terminals already booked. In China, the third-generation Beidou Navigation Satellite System has significant applications in fields like transportation, marine fisheries, and emergency rescue. The remote monitoring system for power transmission facilities in the Yulin Substation of Guangxi is a pilot application based on the Beidou Navigation Satellite System, effectively saving manpower, improving efficiency, and reducing costs. The Hongyan Satellite System, under construction by China Aerospace Science and Technology Corporation, plans to collect data information through a constellation of 60 low-orbit small satellites, achieving all-weather, all-time, global coverage for two-way communication. The "Xingyun Project" by China Aerospace Science and Industry Corporation aims to build a satellite IoT platform using 80 low-orbit satellites to provide communication services for global IoT terminals.
	
	\section{Technical Characteristics and Architecture of Satellite IoT}
	
	\subsection{Technical Characteristics}
	
	The technical characteristics of satellite IoT are mainly reflected in the following aspects:
	
	\begin{itemize}
		\item \textbf{Wide Coverage and Large Capacity}: The use of satellite communication can overcome the limitations of ground networks, providing global coverage, especially in remote, mountainous, and ocean areas where ground communication infrastructure is lacking.
		\item \textbf{High Reliability and Strong Disaster Resistance}: Satellite communication is less affected by natural disasters and can maintain communication even in extreme conditions such as earthquakes, floods, and fires.
		\item \textbf{Flexible Networking and Rapid Deployment}: Satellite IoT can provide flexible networking options and be quickly deployed without being constrained by ground infrastructure construction.
		\item \textbf{Low Power Consumption and Long Battery Life}: IoT devices in satellite IoT systems can achieve low power consumption and long battery life through optimized
   design and efficient communication protocols.
   \item \textbf{Secure and Private Communication}: Satellite IoT can provide secure and private communication services through encryption and authentication mechanisms, ensuring data security and user privacy.
\end{itemize}

\subsection{Architecture of Satellite IoT}

\begin{figure}[h]
\centering
\includegraphics[width=0.9\textwidth]{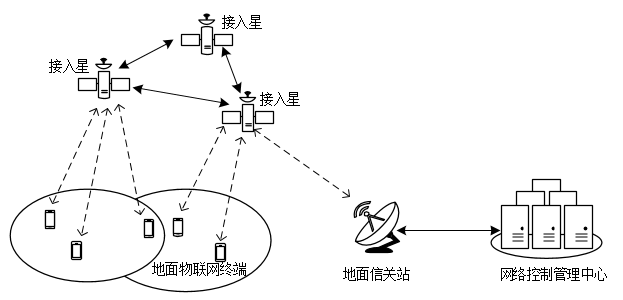}
\caption{Architecture of Satellite IoT}
\label{fig:satellite_iot_architecture}
\end{figure}

The architecture of satellite IoT mainly includes three layers: the space layer, the ground layer, and the user layer, as shown in Figure \ref{fig:satellite_iot_architecture}.

\begin{itemize}
\item \textbf{Space Layer}: The space layer mainly consists of low earth orbit (LEO) satellites, medium earth orbit (MEO) satellites, and geostationary earth orbit (GEO) satellites. These satellites form a constellation to provide global coverage and communication services.
\item \textbf{Ground Layer}: The ground layer mainly consists of gateway stations, monitoring and control centers, and network management centers. The gateway stations are responsible for the communication between the space layer and the ground layer, converting satellite signals into terrestrial network signals. The monitoring and control centers are responsible for the operation, monitoring, and control of the satellites, ensuring the normal operation of the satellite constellation. The network management centers are responsible for the management of the entire satellite IoT network, including resource allocation, performance monitoring, and fault management.
\item \textbf{User Layer}: The user layer mainly consists of various IoT terminals and user equipment. IoT terminals include sensors, actuators, and communication modules, which collect and transmit data. User equipment includes mobile phones, computers, and other devices that access the satellite IoT network through the gateway stations.
\end{itemize}

\begin{figure}[h]
	\centering
	\includegraphics[width=0.9\textwidth]{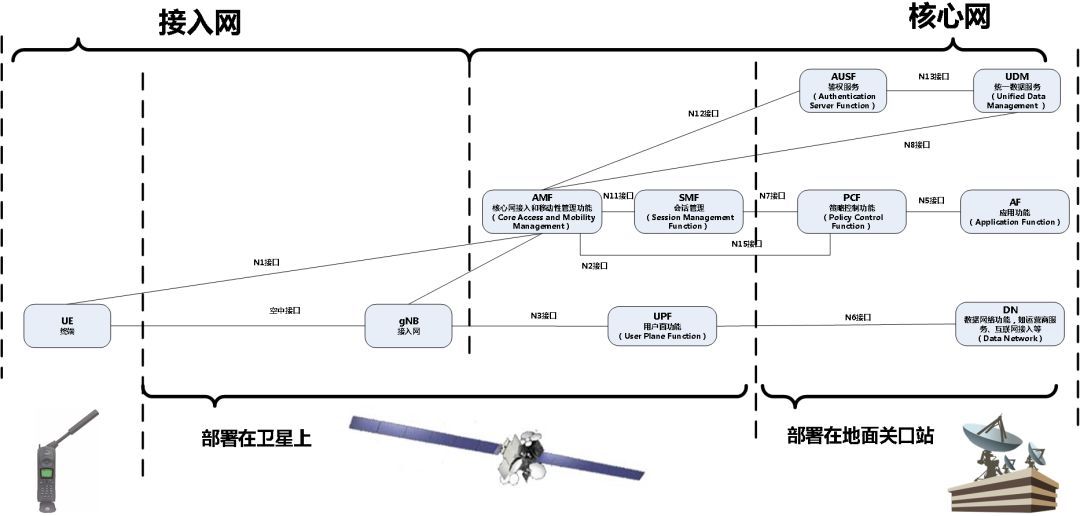}
	\caption{The integrated architecture of satellite communication and 5G system \cite{8766619,8442436,8473415}.}
	\label{fig:integrated}
\end{figure}

\section{Key Technologies of Satellite IoT}

\subsection{Satellite Communication Technology}

Satellite communication technology is the foundation of satellite IoT, mainly including satellite communication systems, satellite transponders, satellite antennas, and satellite communication protocols.

\begin{itemize}
\item \textbf{Satellite Communication Systems}: Satellite communication systems include space subsystems, ground stations, and user terminals. The space subsystem consists of communication satellites and their payloads, which receive, amplify, and retransmit signals. The ground stations are responsible for the uplink and downlink of signals, converting satellite signals into terrestrial network signals. The user terminals are devices that communicate with the satellites, such as IoT terminals and user equipment.
\item \textbf{Satellite Transponders}: Satellite transponders are devices that receive, amplify, and retransmit signals in the satellite communication system. They are the core components of communication satellites, determining the communication capacity and coverage of the satellite.
\item \textbf{Satellite Antennas}: Satellite antennas are devices that transmit and receive signals between the satellites and the ground stations. They include reflector antennas, phased array antennas, and microstrip antennas, among others.
\item \textbf{Satellite Communication Protocols}: Satellite communication protocols are the rules and standards for communication between satellites, ground stations, and user terminals. They include frequency allocation, modulation and coding schemes, and communication procedures.
\end{itemize}

\subsection{IoT Communication Technology}

IoT communication technology is the key to realizing the connection and communication between IoT devices, mainly including short-range communication technology, long-range communication technology, and low-power wide-area network (LPWAN) technology.

\begin{itemize}
\item \textbf{Short-Range Communication Technology}: Short-range communication technology includes Wi-Fi, Bluetooth, Zigbee, and other technologies, which are suitable for short-distance and high-speed data transmission.
\item \textbf{Long-Range Communication Technology}: Long-range communication technology includes cellular communication, satellite communication, and other technologies, which are suitable for long-distance and large-area data transmission.
\item \textbf{Low-Power Wide-Area Network (LPWAN) Technology}: LPWAN technology includes NB-IoT, LoRa, Sigfox, and other technologies, which are suitable for long-distance, low-power, and large-scale IoT applications.
\end{itemize}

\subsection{Edge Computing and Cloud Computing Technology}

Edge computing and cloud computing technology are important support technologies for satellite IoT, mainly including data processing, storage, and analysis.

\begin{itemize}
\item \textbf{Edge Computing}: Edge computing refers to the processing and analysis of data at the edge of the network, close to the source of the data. It can reduce the latency and bandwidth consumption of data transmission, improve the real-time performance and efficiency of data processing, and enhance the security and privacy of data.
\item \textbf{Cloud Computing}: Cloud computing refers to the processing and analysis of data in the cloud, providing scalable, flexible, and cost-effective computing and storage resources. It can support large-scale data processing and analysis, enable remote access and collaboration, and provide advanced analytics and machine learning capabilities.
\end{itemize}

\subsection{Study on Low Latency Network Slicing Technology for IoT}

\begin{figure}[h]
	\centering
	\includegraphics[width=0.9\textwidth]{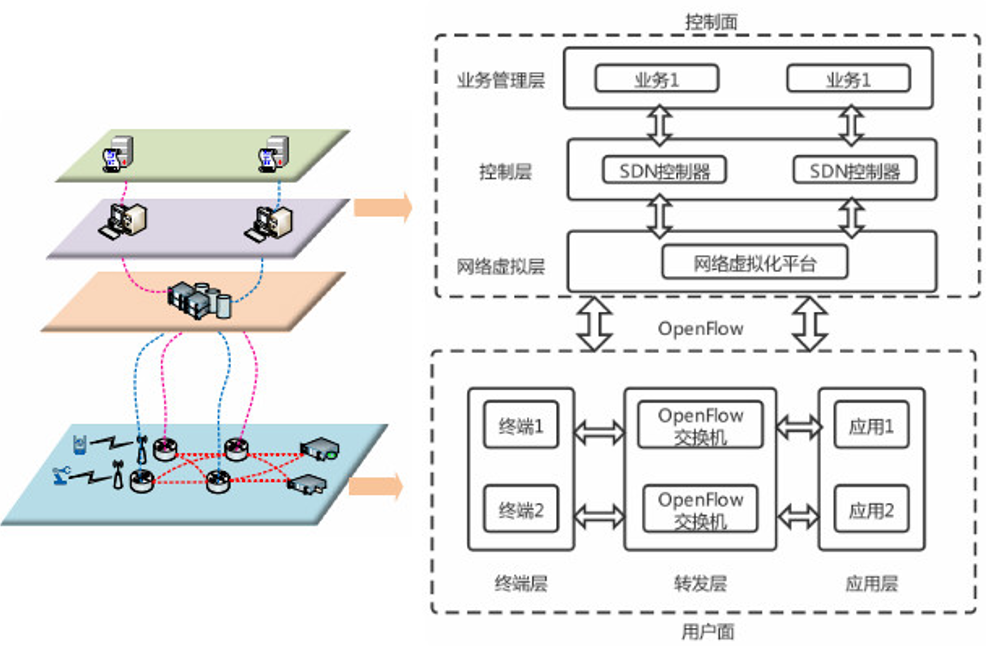}
	\caption{SDN and NFV framework for network slicing.}
	\label{fig:slice}
\end{figure}

\begin{itemize}
	\item In order to provide a low-latency network environment for the deployment of edge computing nodes and handover scenarios, it is necessary to design an efficient 5G network slicing strategy. Research Topic 2 aims to address the issue of building a 5G network slicing topology by dividing network slices from the control plane and user plane based on SDN (Software-Defined Networking) and NFV (Network Functions Virtualization). On this basis, it will manage and orchestrate network slice resources, dynamically adjust path weights using backpropagation neural network theory, and determine the optimal mapping path using heuristic algorithms to resolve latency issues in 5G network slicing.
	\item 2.2.1 Research on Network Slicing Architecture Based on SDN and NFV Technology
	The proposed method draws on the SDN concept of separating data and control planes, dividing network slices from both the control plane and the user plane to facilitate the orchestration and management of user plane network slices by the control plane network slices. Simultaneously, NFV is used to virtualize the network functions of the access network and the core network, achieving software-hardware decoupling, thereby constructing an end-to-end network slicing architecture.
	\item Scientific Issue 2: In IoT networks, the emergence of various types of services makes it crucial to reasonably divide network slices and avoid congestion. How to design an efficient network slicing strategy, build a network architecture that can quickly forward data, manage and orchestrate network slice resources, and find the shortest mapping path for network slices to reduce transmission latency is the second problem this project aims to solve. This project intends to use SDN and NFV technologies to segment and virtualize the network functions of the access network, transport network, and core network, rapidly processing network slices from the user plane and data plane to meet the service demands of different applications. It will also manage and orchestrate network slice resources, using a backpropagation neural network approach to train with large amounts of sample data and dynamically adjust path weights. On this basis, heuristic algorithms will be used to traverse and find the optimal mapping path for network slices, thus improving transmission efficiency and providing a suitable network environment for the deployment of edge computing nodes and handover scenarios.
\end{itemize}

\section{Application Scenarios of Satellite IoT}

Satellite IoT has a wide range of application scenarios, mainly including smart agriculture, smart transportation, smart cities, environmental monitoring, and emergency response.

\subsection{Smart Agriculture}

Satellite IoT can support smart agriculture by providing real-time monitoring and control of agricultural activities, such as soil moisture, crop growth, and pest control. It can enable precision agriculture, optimize resource use, and improve crop yield and quality.

\subsection{Smart Transportation}

Satellite IoT can support smart transportation by providing real-time monitoring and management of transportation systems, such as vehicle location, traffic conditions, and road safety. It can enable intelligent transportation systems, reduce congestion and accidents, and improve transportation efficiency and safety.

\subsection{Smart Cities}

Satellite IoT can support smart cities by providing real-time monitoring and management of urban infrastructure and services, such as energy, water, waste, and public safety. It can enable smart city applications, improve resource use, and enhance the quality of life for residents.

\subsection{Environmental Monitoring}

Satellite IoT can support environmental monitoring by providing real-time data on environmental conditions, such as air quality, water quality, and weather. It can enable early warning systems, support environmental protection, and enhance disaster response and resilience.

\subsection{Emergency Response}

Satellite IoT can support emergency response by providing real-time communication and coordination in disaster situations, such as earthquakes, floods, and fires. It can enable rapid response, improve situational awareness, and enhance rescue and recovery efforts.

\section{Conclusion}

Satellite IoT is an emerging technology that combines the advantages of satellite communication and IoT, providing global coverage, high reliability, and flexible networking. It has a wide range of applications in various fields, including smart agriculture, smart transportation, smart cities, environmental monitoring, and emergency response. With the continuous development of satellite communication, IoT, edge computing, cloud computing, AI, and ML technology, satellite IoT will play an increasingly important role in the future, supporting the digital transformation and sustainable development of society.

\bibliographystyle{IEEEtran}
\bibliography{reference}
\vfill

\end{document}